\documentclass[12pt]{iopart}

\def\be{\begin{equation}}
\def\te{\end{equation}}
\def\ee{\end{equation}}
\def\ba{\begin{eqnarray}}
\def\bea{\begin{eqnarray}}
\def\nn{\nonumber\\}
\def\tea{\end{eqnarray}}
\def\ea{\end{eqnarray}}
\def\eea{\end{eqnarray}}
\def\x{\mathbf{x}}
\def\y{\mathbf{y}}
\def\z{\mathbf{z}}
\def\p{\mathbf{p}}
\def\e{\epsilon}

\def\d{\delta}

\def\L{\Lambda} 
\def\w{\omega}
\def\de{\delta\epsilon}

\usepackage{graphicx}
\usepackage[english]{babel} 
\usepackage[utf8]{inputenc}
\begin{document}
\title{A functional renormalization method for wave propagation in random media}
\author{Federico Lamagna and Esteban Calzetta}
\address{ Physics Department, Buenos Aires University, Ciudad Universitaria, Pabell\'on I, Buenos Aires, 1428, Argentina and IFIBA-CONICET}
\eads{\mailto{fede.lamagna@gmail.com}, \mailto{calzetta@df.uba.ar}}

\begin{abstract}
{We develop the exact renormalization group approach as a way to evaluate the effective speed of propagation of a scalar wave in a medium with random inhomogeneities. We use the Martin-Siggia-Rose formalism to translate the problem into a non equilibrium field theory one, and then consider a sequence of models with a progressively lower infrared cutoff; in the limit where the cutoff is removed we recover the problem of interest. As a test of the formalism, we compute the effective dielectric constant of an homogeneous medium interspersed with randomly located, interpenetrating bubbles. Already a  simple approximation to the renormalization group equations turns out to be equivalent to a self-consistent two-loops evaluation of the effective dielectric constant.
}
\end{abstract}
%\noindent \today

\section{Introduction}
The goal of this paper is to implement the renormalization group method \cite{WilKog74,Gol92,Fis98,McC04} as a tool to study wave propagation in disordered media \cite{Bou62,Frisch68,Sheng06}. We take as paradigmatic problem the Helmholtz equation in three dimensions with a stochastic space dependent wave speed
\be 
\left[ \Delta + \omega^2 \epsilon(\mathbf{x}) \right] \phi(\mathbf{x}) = - j(\mathbf{x})
\label{eq1}
\te 
where $\epsilon$ is the inverse square speed of sound of the medium, that is, the dispersion relation reads $k^2 = \epsilon \omega^2$. $\epsilon$ is in turn a stochastic real variable, that can be split in its mean value plus fluctuations:
\be 
\epsilon(\mathbf{x}) = \bar{\epsilon} + \delta\epsilon(\mathbf{x})
\te 
The properties of the medium are encoded into the self correlation $\langle \delta\epsilon(\mathbf{x}) \delta\epsilon(\mathbf{x'}) \rangle = C(\mathbf{x},\mathbf{x'})$, which we will take to be the only nontrivial moment (for the generalization to non-Gaussian processes see \cite{CalFra15}). As a test problem we shall focus on the mean wave propagation, obtained as an ensemble average over all realizations of the noise. This is described by the mean Green function $G = \langle G_\epsilon \rangle$, where $G_\epsilon$ is the Green function for each realization of the noise $\epsilon$. $G$ obeys the equation
\be 
\left[ \Delta + \omega^2 \bar{\epsilon} \right] G(\mathbf{x,y}) = - \delta(\mathbf{x-y}) - \omega^2 \langle G_{\epsilon}(\mathbf{x,y}) \delta\epsilon(\mathbf{x}) \rangle 
\label{eq2green}
\te 
We define the self energy according to 
\be 
\omega^2 \langle G_{\epsilon}(\mathbf{x,y}) \delta\epsilon(\mathbf{x}) \rangle = \int \! d^dz \  \Sigma(\mathbf{x,z}) G(\mathbf{z,y})
\te 
which, after multiplying equation (\ref{eq2green}) by the bare green function $G_0$ (the inverse of the nonrandom differential operator on the left hand side), leads to a Dyson equation\cite{Frisch68,VolWol80,WolVol82,VolWol92}
\be 
G = G_0 + G_0 \cdot \Sigma \cdot G 
\te 
where the $\cdot$ denotes the matrix multiplication over the continuous index, that is integration over space coordinates. Although $\epsilon$ is real for all realizations of the noise, the self energy will have in general an imaginary part \cite{Eck10}, meaning that the mean wave is losing energy which is scattered by the inhomogeneities \cite{And58,DepDor09,LMDD12,TFD00,Knot12,KnoWel13}. Our goal is to compute this imaginary part, in the limit of zero momentum, using renormalization group methods, and to compare the result to a perturbative evaluation of the same quantity \cite{CalFra15}.

The renormalization group (RG)\cite{WilKog74,Gol92,Fis98,McC04} denotes a cluster of methods to study problems of diverse complexity. The basic idea is to study how the properties of the physical system change when observed at different length scales. This is regulated through a cutoff $\Lambda$. If the theory is expressed through a generating functional, then one seeks an equation describing the dependence of the generating functional with respect to the cutoff; this approach leads to the so-called functional renormalization group (FRG) \cite{WegHou73,Wet93,KBS10,BGW06,BPR11,Bla11,Bla12}. The actual equation is generally rather complex and demands further approximations to yield concrete results \cite{Mor96}.

The generating functional for waves in disordered media is obtained through the Martin-Siggia-Rose \cite{MSR73,DomGia06,Kam11} or closed time-path formalism \cite{Kam11,CalHu08,ZanCal02,ZanCal06}. This approach, which requires the introduction of auxiliary fields, allows us to write down a generating functional for the wave fields as a functional integral. There are two functionals of interest, the generating function for connected Feynman graphs and its Legendre transform, the generating functional for one particle irreducible graphs, also called ``effective action''. We introduce a cutoff in such a way that the effective action reduces to the ``classical'' action, namely, the model without noise, when $\Lambda\to\infty$, while the full theory is recovered at $\Lambda =0$. One can then find a functional differential equation for the effective action as a function of the cutoff, or else parametrize the effective action in terms of an infinite number of couplings and then obtain a hierarchy of equations for them. This hierarchy is then converted into a closed, finite system by truncation.

For the particular case of the self energy at zero momentum the FRG yields an equation with the structure of a diffusion equation with a source term. We apply the method to a medium made of overlapping spheres placed at random over a homogeneous background \cite{Tor02,CalFra15}. For this particular case we both solve the RG equation numerically and provide an approximate analytic solution, and then compare the result with a self consistent two-loop evaluation of the self energy.

We show that even under strong simplifying assumptions the RG yields a result of whose accuracy compares well with the two loops evaluation. Most importantly, because we implement the RG in such a way that for any value in the cut-off we have a consistent theory, it is possible to implement the RG in such a way that relevant symmetries are explicitly enforced. We believe this makes the RG an important tool which will provide valuable insights in the study of wave propagation in random media.

The paper is organized as follows. In next Section we introduce the basics of the functional formalism. In Section 3 we implement the FRG. In Section 4 we discuss the solutions of the FRG equation for the self-energy. and in Section 5 we apply this formalism to the medium composed of overlapping spheres. We conclude with some brief final remarks on future directions.

\section{The functional formalism} 

There is a way to define a generating functional that can be used to derive all moments of the stochastic field $\phi$ \cite{ZJ93}. Namely,
\be
e^{i W_\phi[J_\phi,J_\phi^*]} = \int \! \mathcal{D}\phi \, \mathcal{D}\phi^* \ P\left[ \phi, \phi^* \right] e^{i (J_\phi \cdot \phi + J_\phi^* \cdot \phi^*)}
\te 

The probability density $P[\phi,\phi^*]$ is derived from the probability density for the stochastic field $\epsilon$, as follows
\be 
P\left[\phi,\phi^*\right] = \int \! \mathcal{D}\epsilon \ Q[\epsilon] \ \delta\left( \phi - \phi[\epsilon,j] \right) \, \delta\left( \phi^* - \phi^*[\epsilon,j^*] \right) 
\te 
where $\phi[\epsilon,j]$ is the solution to the differential equation (\ref{eq1}) for given realization of the noise $\epsilon$ and source $j$, and $Q\left[ \epsilon \right]$ is the probability density for the noise. Using shorthand for the differential operator $\mathbf{D}_\epsilon \equiv \Delta + \omega^2 \epsilon(\mathbf{x})$, we can use the following property of the functional delta  
\be 
\delta \left( \mathbf{D}_\epsilon \phi + j \right) = \frac{\delta\left( \phi - \phi[\epsilon,j] \right)}{\mathrm{det}\left( \mathbf{D}_\epsilon \right)}
\te 
where the same is used with the conjugated field.

Lastly, we can resort to a functional Fourier transform to write 

\be 
\delta \left( \mathbf{D}_\epsilon \phi + j \right) = \int \! \mathcal{D}\psi^* \ e^{i \psi^* \cdot \left[\mathbf{D}_\epsilon \phi + j \right]}
\te 
and to the introduction of ghosts fields for the functional determinant \cite{BalNie95,GAB12a,GAB12b}

\be 
\mathrm{det}\left(\mathbf{D}_\epsilon \right) = \int \! \mathcal{D}c^* \, \mathcal{D}c \ e^{i\, c^* \cdot \mathbf{D}_\epsilon \cdot c}
\te 

These last two steps will be repeated for the conjugated field $\phi^*$, and thus including additional fields $\left\lbrace \psi , d^*, d \right\rbrace$.
Regarding the noise probability, we will take it to be Gaussian, therefore 

\be 
Q\left[ \epsilon \right] = \mathrm{exp}\left[-\frac12 \int \! d^dx \, d^dx' \ \left(\epsilon(\mathbf{x}) - \bar{\epsilon} \right) C^{-1}(\mathbf{x,x}') \left(\epsilon(\mathbf{x}') - \bar{\epsilon} \right) \right]
\te 
This whole procedure allows us to write a generating functional as a path integral of a single ``classical'' action of the fields $\mathcal{X}^\alpha = \left\lbrace \epsilon, \phi , \psi^* , \phi^* , \psi , c^* , c , d^* , d \right\rbrace $. This generating functional will allow us to derive moments of all fields. For this reason, sources $\mathcal{J}^\alpha$ have to be included for each one of them. We can write, then

\be 
e^{i W \left[ \mathcal{J}^\alpha \right]} = \int \! \mathcal{D}\mathcal{X}^N \ e^{i \left( S\left[ \mathcal{X}^\alpha \right] + \mathcal{J}_\alpha \mathcal{X}^\alpha \right) } 
\label{eq4genfun}
\te  
This action takes the form
\be 
\!\!\!\!\!\!\!\!\!\!\!\!\!\!\!\!\!\!\!\!\!\! S\left[\mathcal{X}^\alpha \right] = \frac{i}{2} \delta\epsilon \cdot C^{-1} \cdot \delta\epsilon + \psi^* \cdot \left( \mathbf{D}_\epsilon \phi + j \right) +\psi \cdot \left( \mathbf{D}_\epsilon \phi^* + j^* \right) + c^* \cdot \mathbf{D}_\epsilon \cdot c + d^* \cdot \mathbf{D}_\epsilon \cdot d 
\label{eq14S}
\te

\indent We shall make a distinction between ``matter fields'' $\left\lbrace \phi, \psi^* , \phi^*, \psi \right\rbrace$ and ghost fields $\left\lbrace c^*, c,d^*,d \right\rbrace$. It can be shown that the only role of ghost fields is to enforce that $\langle \delta\e(\mathbf{x}) \delta \epsilon(\mathbf{x}') \rangle = C(\mathbf{x},\mathbf{x}')$ to all orders in perturbation theory, because for each loop of matter fields there is a corresponding loop of ghost fields that, being anti-commuting, carries the opposite sign. Taking this into account, they can be ignored, concentrating on matter fields only. 

It is useful to resort to the effective action, which is a functional of the mean fields and the equations of motion are derived by differentiation. First we will abbreviate notation as follows:

\be \left\{ \begin{array}{cc} 
\mathcal{X^\alpha} \equiv& \left\lbrace \epsilon , \phi^a \right\rbrace \nn
J \equiv& \mathrm{source} \; \mathrm{for}\; \epsilon \nn
J^a \equiv& \mathrm{sources} \; \mathrm{for} \; \mathrm{fields} \; \phi^a \nn
S \equiv& S_\epsilon [\epsilon] + \Delta S [\epsilon, \phi^a]
\end{array} \right. 
\te 
Then, the mean fields are defined as functional derivatives of $W[J,J^a]$: 
\bea 
E &=&\frac{\delta W}{\delta J}=e^{-iW\left[ J,J_a\right] }\int\;\mathcal{D}\epsilon \, \mathcal{D}\phi^a\;\epsilon\;e^{i\left[S + J\epsilon+J_a\phi^a\right]  }\nn  \\
\Phi^a &=&\frac{\delta W}{\delta J_a}=e^{-iW\left[ J,J_a\right] }\int\;\mathcal{D}\epsilon \, \mathcal{D}\phi^a\;\phi^a\;e^{i\left[S+J\epsilon+J_a\phi^a\right]  }
\label{mean}
\tea
Taking a Legendre transform of $W$ we define the effective action
\be 
\Gamma \left[ E,\Phi^a\right] = W \left[ J,J_a\right] - J \cdot E - J_a \cdot \Phi^a
\label{eq5legendre}
\te 
where $J$ and $J_a$ are the values of the sources for given $E$ and $\Phi^a$.

\section{The FRG flow equation}
We will apply the functional renormalization group to evaluate the 1PI effective action. For this, we introduce a cutoff scale $\Lambda$, through which we modify the classical action by replacing the two-point correlation $C$ with $C_\Lambda$, with the fundamental property \cite{BGW06,BPR11,Bla11,Bla12}
\be  
C_\Lambda = \left\{ \begin{array}{cc}
0 & \quad \mathrm{if}\;  \Lambda \to \infty \nn
C & \quad \mathrm{if}\;  \Lambda \to 0
\end{array} \right. 
\te 
With this correlation we explore theories with different fluctuations, from zero fluctuations when $\L = \infty$, to the full correlation when $\L$ is set to zero. 
Defining the generating functional in the same way but with the modified action $S_\Lambda$, we then take the Legendre transform to define the effective average action $\Gamma_\Lambda$. Defining $S_{0,\L} \equiv \frac12 \de \cdot C^{-1}_\L \cdot \de $, with $S_\L = S_{0,\L} + S_i$, this functional is such that
\be 
\lim_{\L \to \infty} \big( \Gamma_\Lambda[E,\Phi^a] - S_{0,\L}[E] \big) \to S_i[E,\Phi^a]
\te 
And taking $\L$ to zero is such that the full effective action is recovered, therefore, $\Gamma_\Lambda$ interpolates between the classical action and the full effective action. We are thus interested in the differential equation that describes the evolution of $\Gamma_\L$, usually called the flow equation.  

One way of performing such cutoff procedure is through a multiplicative function in momentum space, for instance \cite{KBS10}
\be 
\widetilde{C}_\Lambda(\mathbf{k}) = \mathcal{D}\left( \frac{k^2}{\Lambda^2} \right) \, \widetilde{C}(\mathbf{k})
\te 
with the function $\mathcal{D}$ such that 
\be 
\mathcal{D}(x) = \left\{ \begin{array}{cc} 
0 &\quad \mathrm{if}\;  x \ll 1 \nn
1 &\quad \mathrm{if}\;  x \gg 1
\end{array}\right. 
\te  
While the most obvious choice is a step function, in some cases a smooth cutoff is preferred. 

Then, as $\Gamma_\Lambda$ is a Legendre transform of $W_\Lambda$, but the variable $\Lambda$ is untouched by the transformation, the partial derivatives with respect to this parameter coincide
\be 
\frac{\partial}{\partial\Lambda} \Gamma_\Lambda \Big\vert_{E,\Phi^a} =  \frac{\partial}{\partial\Lambda} W_\Lambda \Big\vert_{J,J^a} 
\te 
Taking this derivative from the path integral representation of $W_\Lambda$, we get
\be 
\!\!\!\!\!\!\!\!\!\!\!\!\!\!\!\!\!\!\!\!\!\!\!\!    \frac{\partial W_\Lambda }{\partial\Lambda}  \Big\vert_{J,J^a} = e^{-i W_\Lambda[J,J_a]} \int\! d^dx\,  d^dx' \ \frac{i}{2} \partial_\Lambda C^{-1}_\Lambda(\mathbf{x,x}') \cdot \int \! \mathcal{D}\epsilon \,\mathcal{D}\phi^a \ \delta\epsilon(\mathbf{x}) \delta\epsilon(\mathbf{x'}) e^{i\left(S_\Lambda + J \epsilon + J_a \phi^a \right)} 
\label{eq6partialW}
\te 
In turn, if we differentiate the definition of the mean field E with respect to its source we obtain
\be 
\frac{\delta E(\x)}{\delta J(\y)} = e^{-iW_\Lambda} \int\!\mathcal{D}\epsilon\,\mathcal{D}\phi^a\ i \, \epsilon(\x)\, \epsilon(\y)\, e^{i\left(S_\Lambda + J_\alpha X^\alpha \right)} - i E(\x) E(\y) 
\te 
Now, we write out $\epsilon(\x) = \bar{\epsilon} + \delta\e (\x) $, and $E(\x) = \bar{\e} + \langle \delta\e(\x) \rangle $. Thus
\be 
\frac{1}{i} \frac{\delta E(\x)}{\delta J(\y)} =  \langle \delta\e(\x) \delta\e(\y) \rangle - \langle \delta\e(\x) \rangle \, \langle \delta\e(\y) \rangle
\te 
Substituting this into equation (\ref{eq6partialW}), we get the desired equation 

\be 
\frac{\partial \Gamma_\Lambda }{\partial\Lambda}\Big\vert_{E,\Phi^a} = \int\! d^dx\,  d^dx' \ \frac{i}{2} \partial_\Lambda C^{-1}_\Lambda(\mathbf{x,x}') \left\lbrace \frac{1}{i}  \frac{\delta E(\x)}{\delta J(\x')} +  \langle \delta\e(\x) \rangle \, \langle \delta\e(\x') \rangle \right\rbrace 
\te 
Inspecting the second term, we see that if we define $\Gamma_\Lambda = \Gamma_E[E] + \Gamma_\Phi [E,\Phi^a]$, with
\be 
\Gamma_E \left[ E\right] = \frac{i}{2} \int\! d^dx \, d^dx' \ (E(\x) - \bar{\e})\  C_\Lambda^{-1}(\x,\x') (E(\x') - \bar{\e})
\te 
then we get a differential equation for the second term, 

\be 
\frac{\partial \Gamma_{\Phi,\Lambda} }{\partial\Lambda}\Big\vert_{E,\Phi^a} = \int\! d^dx\,  d^dx' \ \frac12 \partial_\Lambda C^{-1}_\Lambda(\mathbf{x,x}')  \frac{\delta E(\x)}{\delta J(\x')} 
\label{eqgammalambda}
\te 
Suppose we are interested in 
\be 
\frac{\partial}{\partial\Lambda}\frac{\partial^2\Gamma_{\Phi,\Lambda}}{\partial\Phi^a\Phi^b}\vert_{E,\Phi^ a=0}
\label{eqgamma2}
\te 
Notice that we take $E$ to be nonzero, while making zero all other mean values. Taking the functional derivatives, and using chain rule we get to 
\bea 
&&\frac{\delta^2}{\delta\Phi^e \delta\Phi^f} \frac{\delta E(\x)}{\delta J(\x')} = - \frac{\delta E(\x)}{\delta J(\x'')} \left\lbrace   \frac{\delta^2 J(\x'')}{\delta\Phi^e \delta\Phi^c} \frac{\delta^2 \Phi^c}{\delta\Phi^f \delta J(\x')} \right. \nn 
&+& \left. \frac{\delta^2 J(\x'')}{\delta\Phi^f \delta\Phi^c} \frac{\delta^2 \Phi^c}{\delta\Phi^e \delta J(\x')} + \frac{\delta^3 J(\x'')}{\delta\Phi^e \delta\Phi^f \delta E(\x''')} \frac{\delta E(\x''')}{\delta J(\x')}  \right\rbrace 
\label{feyn}
\tea  
where from now on, repeated indexes imply both a summation over discrete ones, and integration over continuous ones. That is, $\Phi_a \Phi^a = \sum_{a} \int \! d^dx \, \Phi_a(\x) \Phi^a(\x)$. Rewriting both mean fields and sources as functional derivatives of $W$ and $\Gamma$ respectively, and replacing in equations (\ref{eqgamma2},\ref{eqgammalambda}) we arrive at 

\begin{eqnarray}
\!\!\!\!\!\!\!\!\!\!\!\!\!\!\!\!\!\!\!\!\!\!\!\! \partial_\Lambda \frac{\delta^2 \Gamma_{\Phi,\Lambda}}{\delta \Phi^e \delta\Phi^f} \biggr \vert_{\Phi^a=0} &= \frac12 \partial_\Lambda C_\Lambda^{-1}(\x,\y)  \frac{\delta^2 W_\Lambda}{\delta J(\x) \delta J(\x')} \frac{\delta^2 W_\Lambda}{\delta J(\y) \delta J(\y')} \times \nonumber \\ 
& \times \left[ 2 \frac{\delta^3\Gamma_\Lambda}{\delta\Phi^e \delta\Phi^c \delta E(\x')} \frac{\delta^2 W_\Lambda}{\delta J^c \delta J_{c'}} \frac{\delta^3\Gamma_\Lambda}{\delta\Phi^f \delta\Phi^{c'} \delta E(\y)} + \frac{\delta^4\Gamma_\Lambda}{\delta\Phi^e \delta\Phi^f \delta E(\x') \delta E(\y')} \right] \label{eqFLOW}
\end{eqnarray}

This is the so-called flow equation for the two point effective vertex \cite{BGW06,BPR11,Bla11,Bla12}, and we see it connects $\partial_\Lambda \Gamma^{(2)}$ with the higher order vertices $\Gamma^{(3)}$ and $\Gamma^{(4)}$, so it's actually a whole hierarchy of equations. Being interested in the effective constant for the medium, we propose a certain functional structure for $\Gamma_\Lambda$, as follows:

\be 
\!\!\!\!\!\!\!\!\!\!\!\!\!\!\!\!\!\!\!\!\!\!\!\!  \Gamma_\Lambda[E,\Phi^a] = \Gamma_0(\Lambda) + \Gamma_E[E,\Lambda) + \frac12 \int \! d^3x  I_{ab} \left\lbrace \Phi^a (\mathbf{x})  \nabla^2 \Phi^b(\mathbf{x}) + \omega^2 \epsilon_\Lambda[E;\mathbf{x}) \, \Phi^a (\mathbf{x}) \Phi^b(\mathbf{x}) \right\rbrace
\label{eq32Gamma}
\te 

The matrix $I_{ab}$ in one that connects the fields $\Phi^a$ respecting the structure of the classical action, for instance it has to connect $\langle \psi^* \rangle $ with $\langle \phi \rangle$. Therefore, it shall have the following form 

\be  
I_{ab} = \left( \begin{array}{cccc}
0&  1  & 0 & 0\\
1  &  0 & 0 & 0\\
0 & 0 & 0 & 1 \\ 
0 & 0 &1 & 0 \\
\end{array} \right)
\te 
Also, we consider the functional $\e_\Lambda[E;\x)$ to have an expansion in gradients \cite{Mor96}
\be 
\epsilon_\Lambda [E;\x) = f_\Lambda (E(\x)) + f_i(E(\x)) \partial^i E(\x) + \frac12 f_{ij}(E(\x)) \partial^i E(\x) \partial^j E(\x)... \\
\te 
of which we will keep just the first term. 

We first take the functional derivatives of $\Gamma_\Lambda$ in order to construct the two-point, three-point and four-point effective vertices:
\begin{eqnarray}
\frac{\delta^2 \Gamma_\Lambda}{\delta \Phi^e(\x) \delta \Phi^f (\y)} = I_{ef} \Big[ \nabla_x^2 \delta(\x-\y) + \omega^2 f_\L(E(\x)) \, \delta(\x-\y) \Big] \equiv I_{ef} \gamma^{(2)}(\x,\y)
\label{gam2}
\end{eqnarray} 
\begin{eqnarray} 
\!\!\!\!\!\!  \frac{\delta^3 \Gamma_\Lambda}{\delta \Phi^e(\x) \delta \Phi^f (\y) \delta E(\x')} = I_{ef} \, \omega^2 f_\Lambda'(E(\x)) \delta(\x-\x') \, \delta(\x-\y) \equiv I_{ef} \gamma^{(3)}(\x,\y;\x')
\end{eqnarray}
\begin{eqnarray} 
\frac{\delta^4 \Gamma_\Lambda}{\delta \Phi^e(\x) \delta \Phi^f (\y) \delta E(\x') \delta E(\y')} &= I_{ef} \, \omega^2 f_\Lambda''(E(\x)) \, \delta(\y'-\x)\, \delta(\x-\x') \, \delta(\x-\y) \nn 
&\equiv I_{ef} \gamma^{(4)}(\x,\y;\x',\y')
\end{eqnarray}
These are inserted in equation (\ref{eqFLOW}). We note, first
\be
\frac{\delta^2 W_\Lambda}{\delta J^c(\x) \delta J_{c'}(\x')} \equiv G_{cc'}(\x,\x')
\te 
And, as $\frac{\delta^2 \Gamma_\Lambda}{\delta \Phi^e(\x) \delta \Phi^f (\y)}$ is proportional to the matrix $I_{ef}$, which is its own inverse. Then $G_{cc'}$ is proportional to the same matrix, such that 
\be 
\sum_f \, \int\! d^dx' \, \frac{\delta^2 \Gamma_\Lambda}{\delta \Phi^e(\x) \delta \Phi^f (\x')} \frac{\delta^2 W_\Lambda}{\delta J_f(\x') \delta J_{g}(\y')} = - \delta_e^g \, \delta(\x,\y')
\te 
\noindent a property that derives from the Legendre transform. This allows us to write $\Gamma^{(2)}_{ef} \equiv I_{ef} \gamma^{(2)}$, and $G_{cc'} \equiv I_{cc'}g$, because the $I$ matrix is its own inverse. We will be interested in the case in which the mean field is homogeneous, $E = \bar{\e}$. Then, 
\be 
\frac{\delta^2 W_\Lambda}{\delta J(\x) \delta J(\x')} = i \left[ C_\L(\x,\x') - \langle \delta\e(\x) \rangle \langle \delta\e(\x') \rangle  \right]
\te 
and we can use
\be 
\int \! d^dx' d^dy' \  C_\L (\x,\x') \partial_\L C^{-1}_\L(\x',\y') C_\L(\y',\y) = - \partial_\L C_\L (\x,\y)
\te  
We thus arrive at 
\be 
\!\!\!\!\!\!\!\!\!\!\!\!  \partial_\Lambda \gamma^{(2)}(\x,\y) = \frac12 \dot{C}_\Lambda(\x',\y') \left[ 2 g(\z,\z') \gamma^{(3)}(\x,\z;\x') \gamma^{(3)}(\y,\z';\y') + \gamma^{(4)}(\x,\y;\x',\y')\right]
\te 

Integrating out the term with the four-point vertex, we notice that 
\be
\int \! d^dx' \, d^dy' \ \dot{C}_\Lambda(\x',\y') \,  \gamma^{(4)}(\x,\y;\x',\y') = \dot{C}_\L(0) \, \frac{\partial^2 \gamma^{(2)}}{\partial E^2} (\x,\y)
\te 

\noindent where we are assuming from now onwards that the correlation $C_\Lambda$ is translationally invariant, $C_\L(\x-\y)$. With $\gamma^{(2)}$ also being invariant, the green function $g$ has to be as well. This allows us to perform a Fourier transform of the whole equation. The flow equation for $f(\L,E)$ is obtained by setting $\mathbf{p}=0$ on both sides. This differential equation has to be solved through $\L = 0$, from its initial condition at $\L = \infty$. By requiring that $\Gamma_{\L = \infty} = S$, comparing eqs (\ref{eq14S}) and (\ref{eq32Gamma}), we have 
\be 
f(\L=\infty, E) = E
\te 
And after setting $\p = 0$ we have the equation
\be 
\left[ \frac{\partial}{\partial\Lambda} - \frac12 \dot{C}_\Lambda(0) \frac{\partial^2}{\partial E^2} \right] f(\Lambda,E) =  \int \! \frac{d^3q}{(2\pi)^3} \, \partial_\Lambda \widetilde{C}_\Lambda(q^2) \widetilde{g}(q^2) \, \omega^2 f'^2  
\label{eqefe}
\te 
We are interested with the value of $f$ because its related to the self energy. Indeed, by writing the equation for the propagator $g$, we have:
\be 
\nabla^2 g(\x) + \omega^2 f(E) g(\x) = -\delta(\x)
\te 
as we know that $g$ is the propagator from equation (\ref{eq2green}), by comparing the two expressions, we get 
\be 
f = \bar{\e} +\frac{1}{\w^2} \widetilde{\Sigma}(p=0)
\te 
Thus, $f$ is identified with an effective inverse square speed of sound, $\e_{\textrm{eff}}$ such that $k^2= \e_{\textrm{eff}} \, \omega^2$. This propagator is then written as (3 dimensions)
\be 
g(r = |\x|) = \frac{ e^{ikr} }{4\pi r} \quad\quad\quad \widetilde{g}(p) = \frac{1}{p^2 - \omega^2 f - i \d} \quad \left( \d \to 0\right)
\te 
As $k$ can be complex, its imaginary part will be responsible for an exponential decay of the wave inside the medium.

\subsection{Correlation function and cutoff}
In order to solve equation (\ref{eqefe}), we must first propose a shape for the cutoff. Choosing a step function in momentum space we can readily calculate the integral.
\be 
\widetilde{C}_\L(\mathbf{p})=\Theta(p - \L) \widetilde{C}(\mathbf{p})
\te 
Also, assuming rotational invariance, the correlation function $\widetilde{C}$ will only depend on the absolute value of momentum. This allows us to write 
\be 
\!\!\!\!\!\!  \dot{C}_\L (0) = \int \! \frac{d^3p}{(2\pi)^3} \ \partial_\L \widetilde{C}_\L(p) = - \frac{1}{2\pi^2} \int_0^\infty \! dp \ p^2 \delta(p - \L) \,  \widetilde{C}(p) = - \frac{1}{2\pi^2}  \L^2 \widetilde{C}(\L^2)
\te 
The right-hand side of equation (\ref{eqefe}) can be similarly integrated, to yield
\be 
\left[ \frac{\partial}{\partial\Lambda} - \frac12 \dot{C}_\Lambda(0) \frac{\partial^2}{\partial E^2} \right] f(\Lambda,E) = \dot{C}_\L(0)\, \widetilde{g}(\L^2) \, \omega^2 f'^2  
\te 
We can define a new ``time'' variable, by setting
\be 
t = \frac{C_\L(0)}{C(0)}
\te 
This maps $\L=\infty$ to $t=0$, and $\L = 0$ to $t=1$, and the inverse map will be called $\L(t)$. The factor $C(0)$, being the original correlation between a point and itself, can be also absorbed by a change in the other variable
\be 
e = \frac{E}{\sqrt{C(0)}}
\te 
Finally, calling $R$ the length of correlation, such that $C(\x)$ will be tipically non-zero inside the sphere $|\x| \leq R$. With this, defining $ X = R \L(t)$, and $F = \w^2 R^2 f$, we reach the adimensional differential equation
\be 
\left[ \frac{\partial}{\partial t} - \frac12 \frac{\partial^2}{\partial e^2} \right] F(t,e) = \frac{1}{X^2(t) - F} \left(\frac{\partial F}{\partial e}\right)^2
\label{Fte}
\te 
with initial condition 
\be 
F(t=0,e) = \w^2 R^2 \sqrt{C(0)} e \equiv Q e
\te 
This parameter $Q$ is called the \textit{generalized Reynolds number}, and is a measure of the strength of randomness in the medium \cite{Frisch68}. 

\section{Parametric solution on imaginary part}

Focusing on the imaginary part of the solution, that will be related to the length of exponential decay of the coherent wave, we write $F = \textrm{Re}(F) + i\sigma$. The initial condition, $F = Qe$, implies $\sigma(t=0,e) =0$, and we will take $\sigma \ll 1$ as well. As $X(t=0) = \infty$, for $t \ll 1$, we have 
\be 
X^2(t) \gg F 
\te  
And therefore we have, writing $\textrm{Re}(F) = Qe + \alpha(t)$, 
\be 
\frac{\partial}{\partial t} \alpha(t) = \frac{Q^2}{X^2(t)}
\te 
Which can be readily integrated to yield 
\be 
F(t \ll 1 ,e) = Qe + Q^2 \, \int_0^t \! \frac{dt'}{X^2(t')}
\te 
Inverting the relationship between $X$ and $t$ we obtain 
\be 
\frac{dt}{dX} = \frac{1}{R} \frac{dt}{d\L} = \frac{1}{R} \frac{\dot{C}_\L(0)}{C(0)} = - \frac{\L^2}{2\pi^2R} \frac{\widetilde{C}(\L)}{C(0)} \equiv - \frac{X^2}{2\pi^2 R^3}  \frac{\widetilde{C}(X)}{C(0)}
\te 
 
For the imaginary part, we first write the right hand side as
\be 
\frac{1}{X^2 - F} = \frac{X^2 - F^*}{|X^2 - F|^2} = \frac{X^2 - \textrm{Re}(F) + i \sigma }{ \left( X^2 - \textrm{Re}(F)\right)^2 + \sigma^2 }
\te 
and 
\be 
\left(F'\right)^2 = \left(Q + i \sigma' \right)^2
\te 
Then, taking both $\sigma \ll 1$ and $\sigma' \ll 1$, we take
\be 
\left(F'\right)^2 \sim Q^2
\te 
Thus we get the equation
\be 
\left[ \frac{\partial}{\partial t} -  \frac12 \frac{\partial^2}{\partial e^2} \right] \sigma(t,e) = Q^2 \, \frac{ \sigma }{ \left( X^2 - \textrm{Re}(F) \right)^2 + \sigma^2 }
\te 
The right hand side, being a Cauchy distribution, when $\sigma \ll 1$ we can write
\be 
\left[ \frac{\partial}{\partial t} - \frac12 \frac{\partial^2}{\partial e^2} \right] \sigma(t,e) = Q^2 \pi \d\left(X^2(t) - \textrm{Re}(F)\right)
\te 
$X$ goes from $\infty$ to $0$ when $t$ goes from 0 to 1. Thus, there is is a value of $t=t^*(e)$ such that the argument of the delta function becomes zero, when 
\be 
\textrm{Re}(F) = X^2(t^*)
\te 
Then, for each $e$, there is a ``time'' $t^*$ at which $\sigma$ develops its final value. This solution at $t=1$ can be written in parametric form, as
\begin{eqnarray} 
e(X) &= \frac{X^2}{Q} - \frac{Q}{2\pi^2 R^3} \int_X^\infty \! dX' \  \frac{\widetilde{C}(\L(X'))}{C(0)} \label{param1} \\ 
\sigma(X) &= \frac{\pi \, Q^2 \, X^2 \, \widetilde{C}(\L(X))}{Q^2 \, \widetilde{C}(\L(X)) + 4 \pi^2 X R^3 C(0)}  \label{param2}
\end{eqnarray}
\section{Effective dielectric constant for overlapping spheres}
In order to check the validity of the method, we use a correlation function for overlapping spheres in a homogeneous background with difference in dielectric constant, as found in \cite{Tor02,CalFra15}. In the limit of low density, we have 
\be
\left\{ \begin{array}{cc}
&\bar{\epsilon} = \e_0 + \Delta \e \, \rho v\\
&C(\x,\x') = \left( \Delta \e \right)^2 \rho v \, \Omega(\x,\x')  \label{correl}
\end{array} \right.
\te
with $\Delta\e$ the difference in the dielectric constant between background and spheres, $\rho$ the number density of spheres, and $v \Omega(\x,\x')$ the volume of intersection between two spheres with centers located in $\x$ and $\x'$. However, as that system of spheres is not actually Gaussian, we are ignoring all higher moments. The advantage of this correlation function is that its momentum space form can be exactly calculated. 
\be 
\widetilde{C}(kR) =  12 \pi R^3  C(0)\, \left[ \frac{\sin(kR) - k R \cos(kR) }{(kR)^3} \right]^2 
\label{ckr}
\te 

\begin{figure}[!htb]
 \includegraphics[scale=0.7]{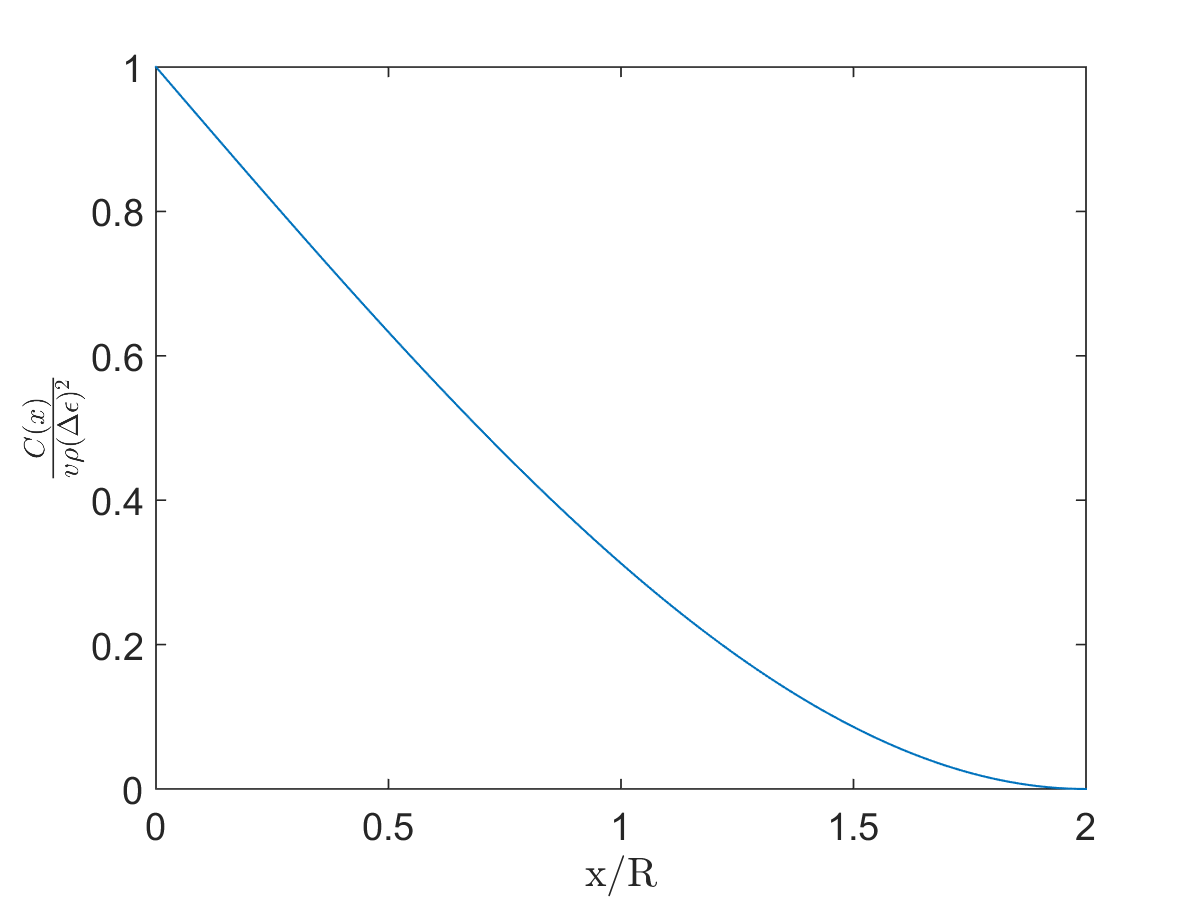}
\caption{Position space correlation function eq. (\ref{correl}). Both the coordinate variable and the correlation function have been rendered dimensionless.}
\label{correlacionx}
\end{figure}

\begin{figure}[h!]
\begin{center}
\includegraphics[height=8cm]{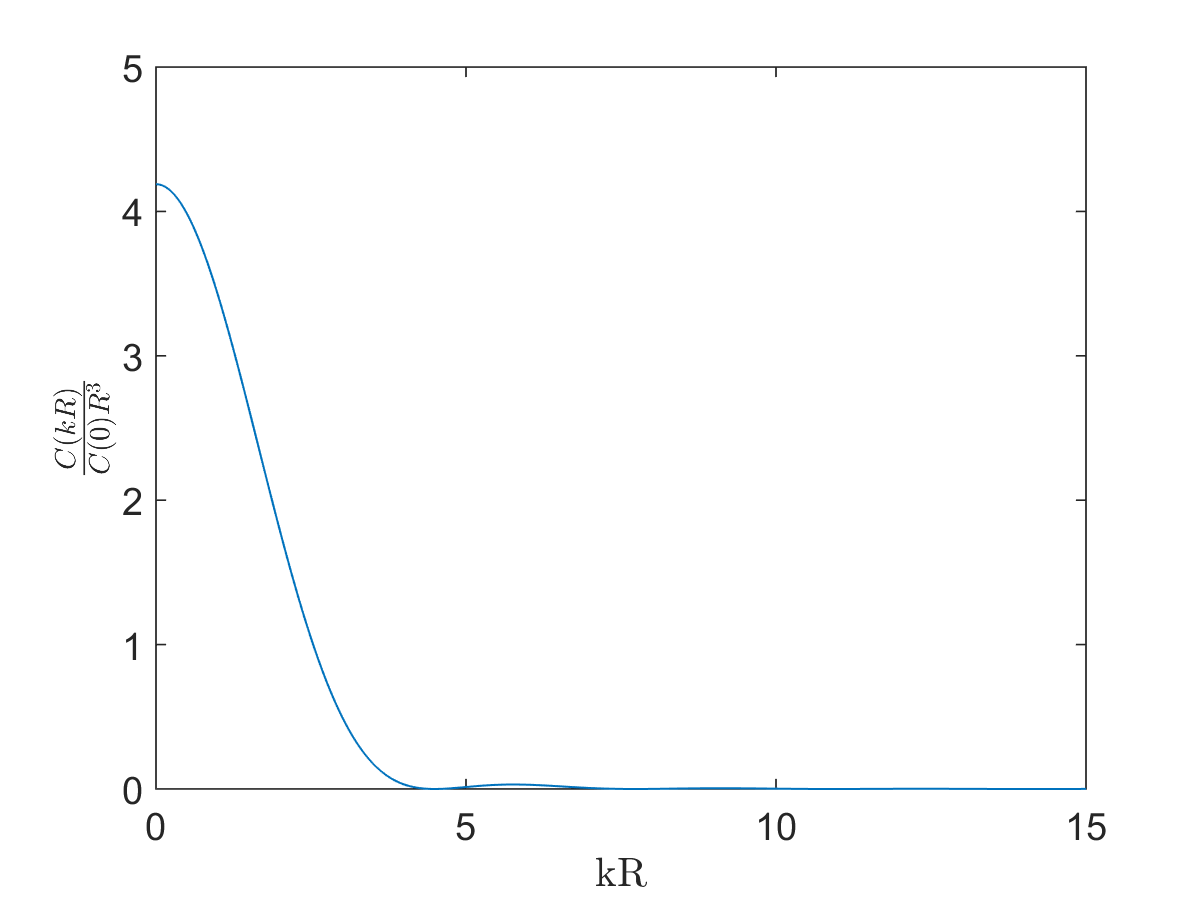}
\caption{Momentum space correlation function eq. (\ref{ckr}). Both the momentum variable and the correlation function have been rendered dimensionless.}
\label{correlimp}
\end{center}
\end{figure} 

This allows for the parametric solution to be obtained readily. As to the full differential equation, the function $t(X)$ can be calculated explicitly, 
\be 
t(X) = \frac{1}{ \pi X^3} \Bigg\lbrace  1 + 3 X^2 - 2 \, \sin(2X) - \cos(2X) \left( 1 + X^2 \right)- 2 X^3 \, \left( \mathrm{sinI}(2X) - \frac{\pi}{2} \right) \Bigg\rbrace
\label{tequis}
\te 

\begin{figure}[ht]
\begin{center}
\includegraphics[height=6cm]{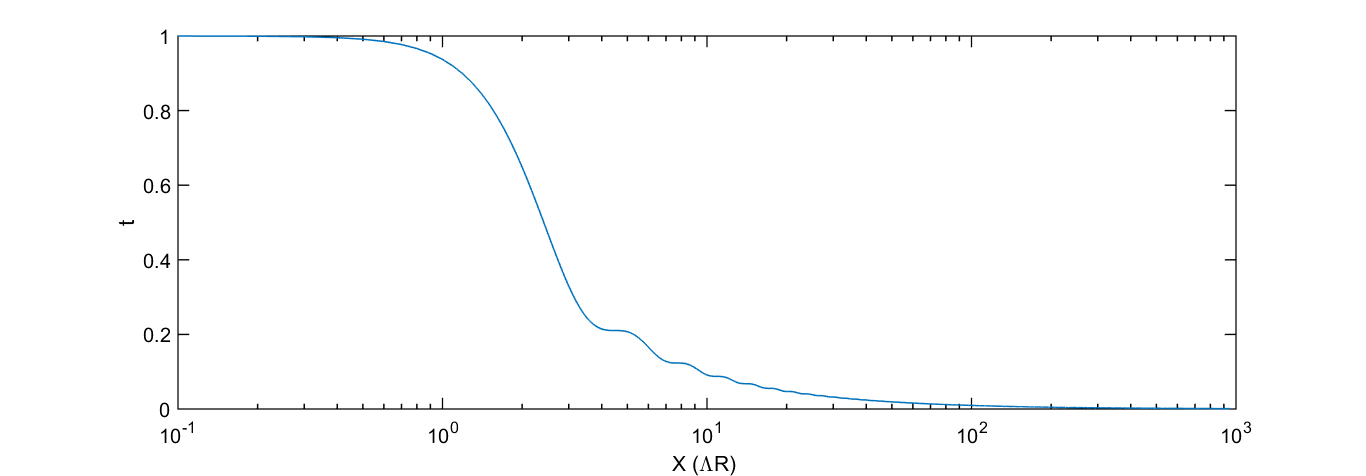}
\caption{Plot of $t(X)$ from eq. (\ref{tequis}).)}
\label{tX}
\end{center}
\end{figure}
\begin{figure}[ht!]
\begin{center}
\includegraphics[height=8cm]{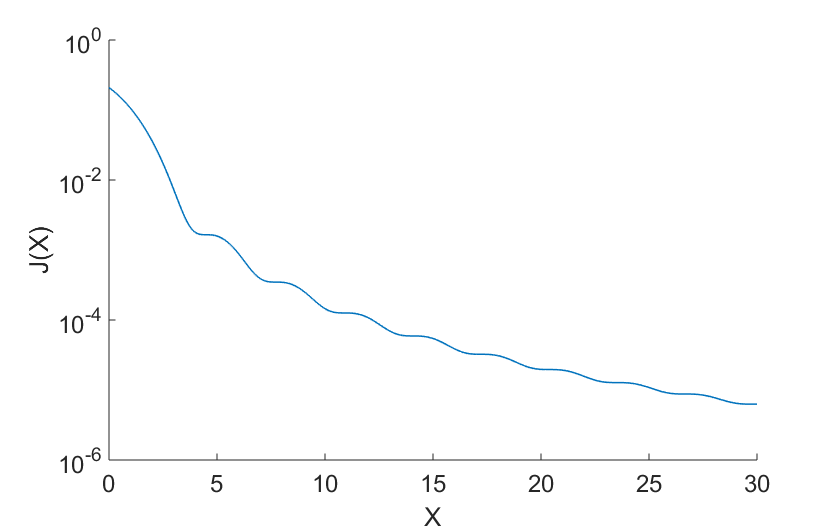}
\caption{Plot $J(X)$ from eq. (\ref{jx}) .}
\label{JX}
\end{center}
\end{figure}

We also obtain the explicit dependence
\be 
\frac{\widetilde{C}(X)}{R^3 C(0)} = \frac{12 \pi \left(\sin X - X \cos X \right)^2}{X^6} 
\te 
The parametric equation becomes
\bea 
e&=& Q^{-1} X^2- \frac{6 Q}{\pi} J(X)  \label{eex} \\
\sigma\left( e\right)&=&\frac{3 Q^2 X^2 \left(\sin X - X \cos X \right)^2 }{ X^7 + \frac{3 Q^2}{\pi} \left(\sin X - X \cos X \right)^2    } \label{sigme} 
\tea
where
\be 
J(X) = \int_X^\infty \! dY \; \frac{\left(\sin Y - Y \cos Y \right)^2}{Y^6} 
\te 
Explicitly
\begin{eqnarray}    
J(X) &= X^{-5} \Bigg\lbrace \frac{1}{10}\left(1 - \cos(2X)  \right)  - \frac15 X \sin(2X) + \frac{1}{30} X^2 \left(\cos(2X) + 5 \right)  \nonumber \\ 
& \quad - \frac{1}{30} X^3 \sin(2X) - \frac{1}{15} X^4 \cos(2X) - \frac{2}{15} X^5 \left( \mathrm{sinI}(2X) - \frac\pi2 \right) \Bigg\rbrace 
\label{jx}
\end{eqnarray}  
which is plotted in fig. (\ref{JX}).

We evaluated numerical solutions to the differential equation (\ref{Fte}) for different values of the parameter $Q$, alongside with parametric solutions to the differential equation. 

\begin{figure}[h!]
\begin{center}
\includegraphics[height=8cm]{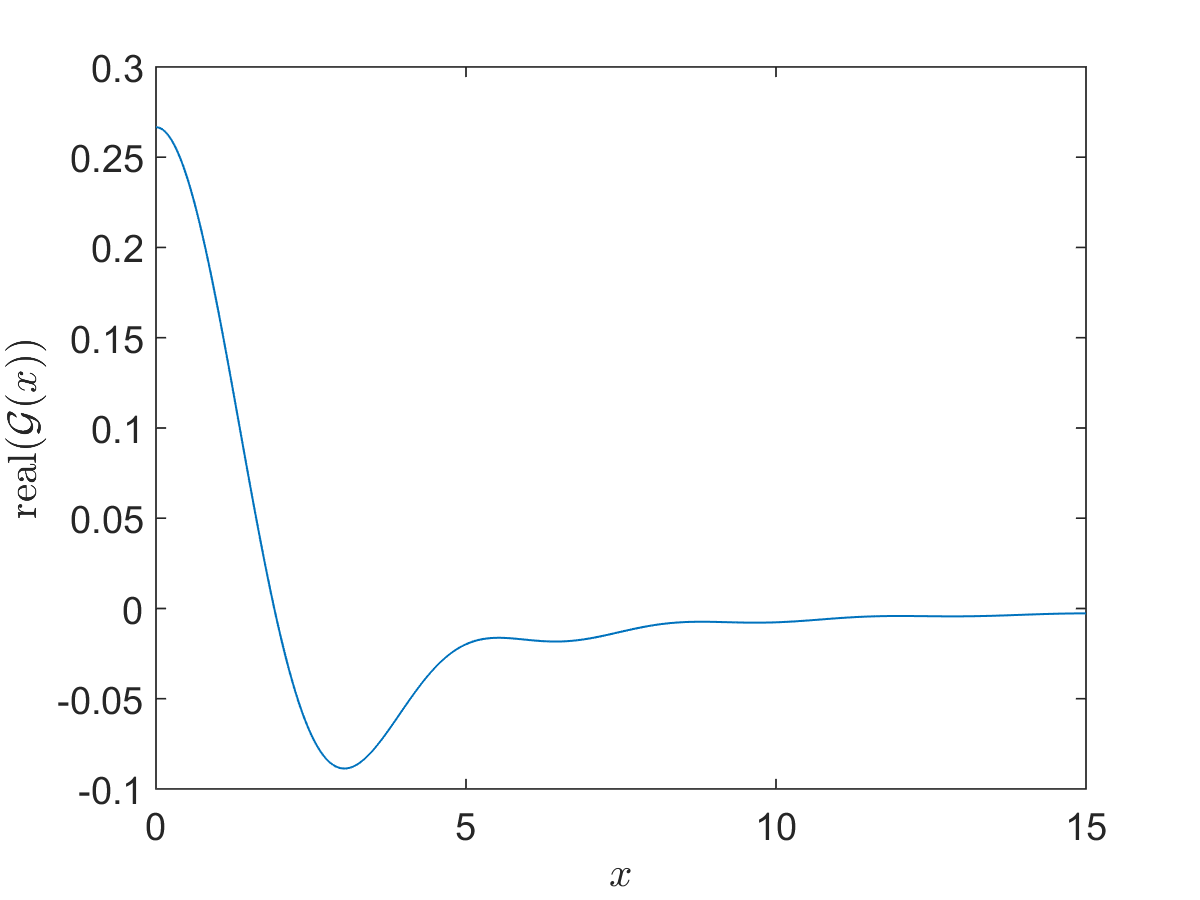}
\caption{Plot of the real part of $\mathcal{G}(x)$ from eq. (\ref{mathcalg}).}
\label{Gxr}
\end{center}
\end{figure}

\begin{figure}[h!]
\begin{center}
\includegraphics[height=8cm]{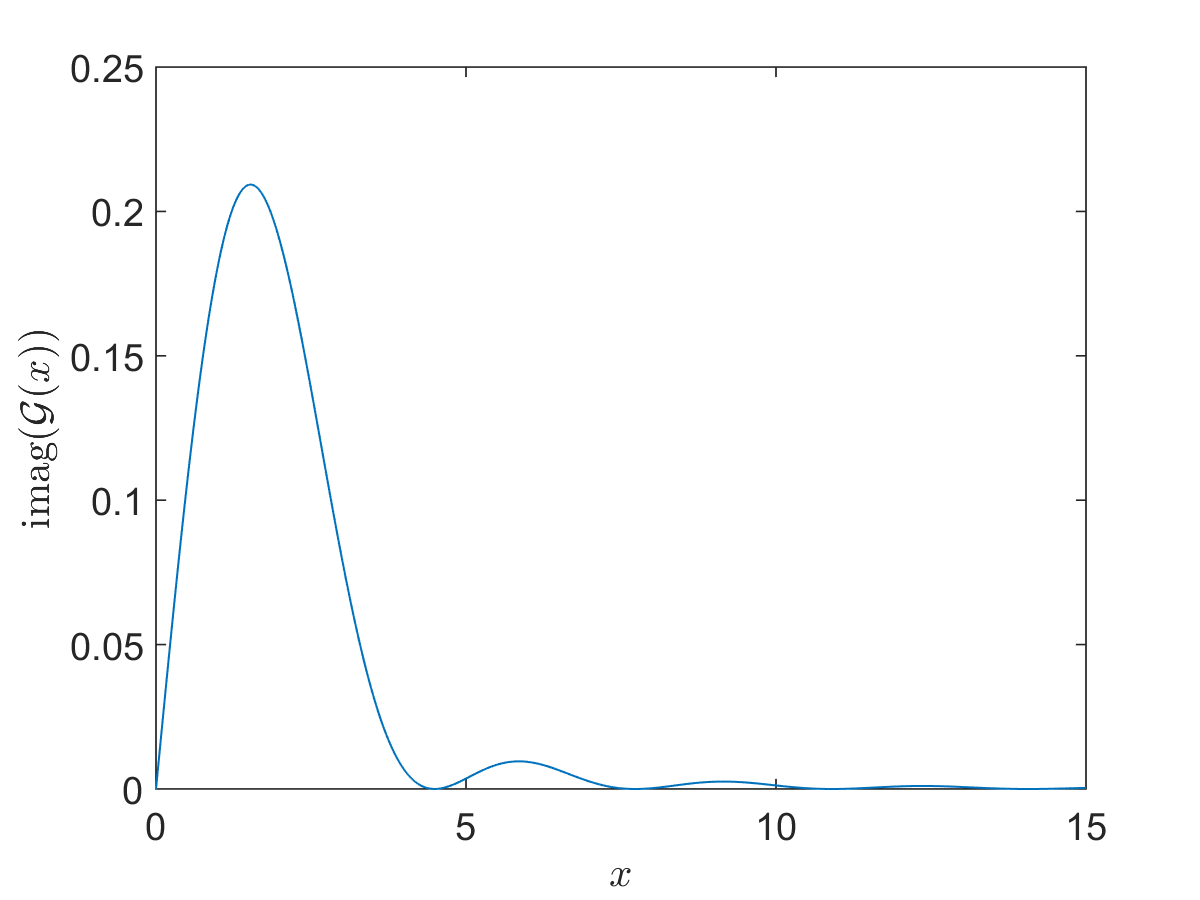}
\caption{Plot of the imaginary part of $\mathcal{G}(x)$ from eq. (\ref{mathcalg}).}
\label{Gxi}
\end{center}
\end{figure}

We also compared these solutions to a self consistent calculation  of the effective dielectric constant from the 1 and 2 loop diagrams in the non-linear approximation to the self energy  \cite{CalFra15}

\be 
\widetilde{\Sigma}(0) = \widetilde{\Sigma}^{(1)}_{NL}(0)+ \widetilde{\Sigma}^{(2)}_{NL}(0)
\te 
At zero momentum this becomes an equation for the effective dielectric constant
\be 
\e_{\textrm{eff}} = \bar{\e} + \e_1 + \e_2
\label{eeff}
\te 
We compute the relevant Feynman graphs approximating the self energy by the effective dielectric constant itself. Then
\be 
\e_1 = \w^2 \Delta\e^2 \rho \, \int\! \frac{d^3q}{(2\pi)^3} \ \frac{v \widetilde{\Omega}(q)}{q^2 - \e_{\textrm{eff}} \w^2}
\te 
Explicitly \cite{CalFra15}
\be 
\e_1 = \frac32 \, \w^2 R^2 \Delta\e^2 \rho v \ \mathcal{G}(\w R\sqrt{\e_{\textrm{eff}}})
\label{epsi1loop}
\te 
Recall $Q =  \w^2 R^2 \sqrt{C(0)}$,  $C(0) = \rho v \Delta\e^2$, and $ F = \w^2 R^2 f$, whereby 
\be 
F_1 = \w^2 R^2 \e_1 = \frac32 \, Q^2 \ \mathcal{G}(\sqrt{Qe + F_1 + F_2}\,)
\label{f1}
\te 
where
\be 
\mathcal{G}(x) = \frac{i}{x^5} \left[ 1 - e^{2ix} + x^2 \left( 1+ e^{2ix}\right) + 2 i x\left( \frac{x^2}{3} + e^{2ix} \right)    \right] 
\label{mathcalg}
\te 
whose real and imaginary parts are plotted in figs. (\ref{Gxr},\ref{Gxi})

\begin{figure}[h!]
\begin{center}
\includegraphics[height=8cm]{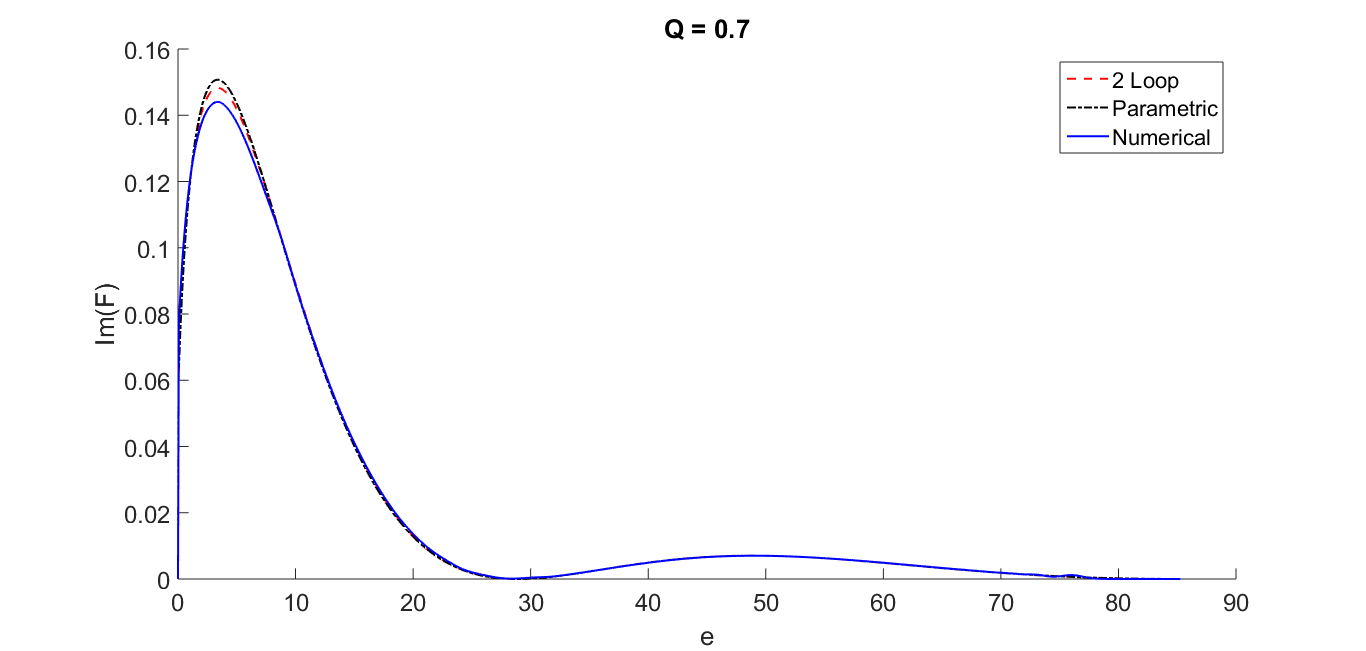}
\caption{The three results concerning the imaginary part of the effective dielectric constant at generalized Reynolds number $Q=0.7$}
\label{QQQQ}
\end{center}
\end{figure}

The two loops contribution is
\be 
\e_2 = \w^6 \Delta\e^2 \rho^2 \, \int\!\frac{d^3q \, d^3p}{(2\pi)^6} \  \frac{v \widetilde{\Omega}(q)v \widetilde{\Omega}(p) }{ \left(p^2-\e_{\textrm{eff}}  \w^2 \right) \left(q^2-\e_{\textrm{eff}}  \w^2 \right) + \left(\left(\p+\mathbf{q}\right)^2-\e_{\textrm{eff}}  \w^2 \right)} 
\te 
whose dominant term is 
\be 
F_2 = \frac{Q^4}{9} \,  \textrm{ln}  \left( \sqrt{Qe + F_1 + F_2}\right) 
\label{f2}
\te 
Using eq. (\ref{f1}) and (\ref{f2}) into (\ref{eeff}) we obtain a self consistent equation for $\epsilon_{eff}$ which may be solved numerically. Figs. (\label{Q}) and (\ref{QQQQ}) show plots of the solution by all three methods (numerical solution of the RGE, parametric approximation to it, and self-consistent two loop calculation) for $Q=1$ and $Q=.7$ respectively. Even at the high value of $Q=1$ results are similar, for $Q$ lower than $0.7$ there are no significant differences.

\begin{figure}[!h]
\begin{center}
\includegraphics[height=8cm]{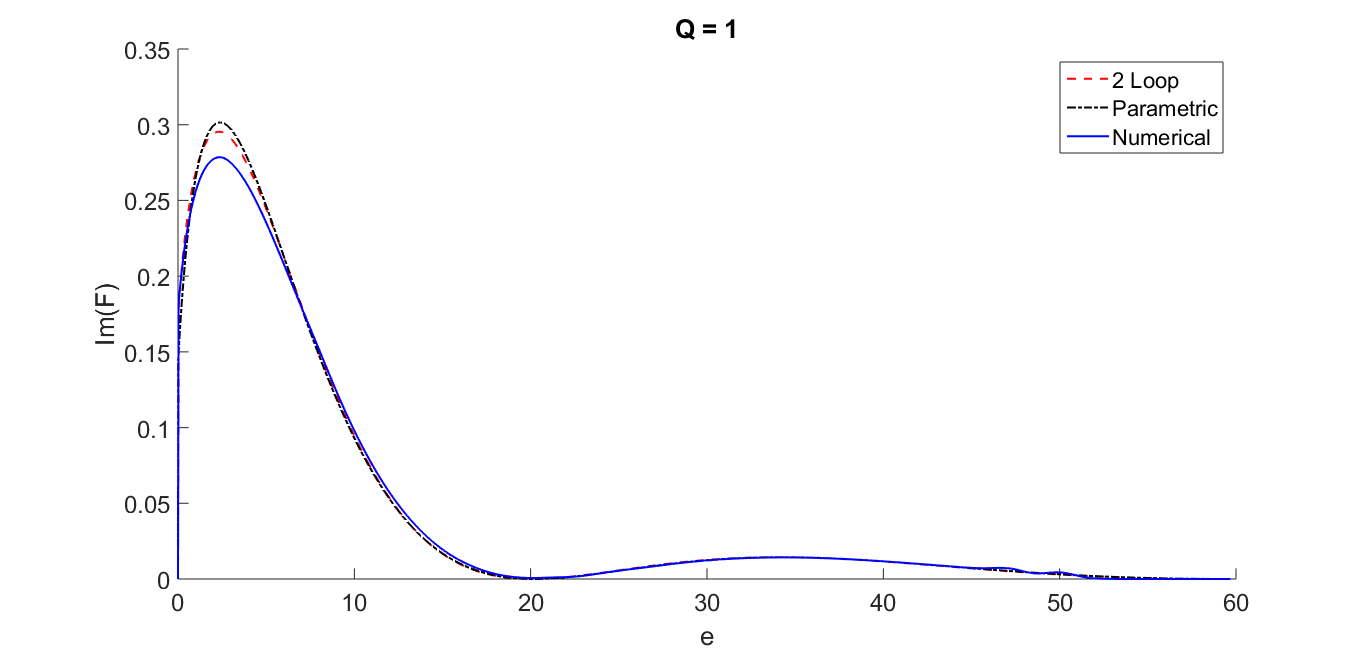}
\caption{The three results concerning the imaginary part of the effective dielectric constant at generalized Reynolds number $Q=1$}
\label{Q}
\end{center}
\end{figure}

\section{Final Remarks}
In this paper we implemented the functional renormalization group as a tool to study wave propagation in random media. This implies three steps. First an effective action for the wave fields is introduced by using the Martin - Siggia - Rose \cite{MSR73,DomGia06,Kam11} or closed time-path formalism\cite{Kam11,CalHu08,ZanCal02,ZanCal06}. Then an infrared cutoff $\Lambda$ is introduced, in such a way that the noise is totally suppressed when $\Lambda\to\infty$, while the original theory is recovered at $\Lambda =0$ \cite{BGW06,BPR11,Bla11,Bla12}. Finally, a renormalization group equation for the effective action is derived by using exact renormalization group techniques \cite{WegHou73,Wet93,KBS10}. Equivalently, we may parametrize the effective action in terms of cutoff-dependent coupling constants, and then the RGE becomes a hierarchy of equations for these constants.

As a demonstration of the method we computed the imaginary part of the self energy at zero momentum for a medium composed of overlapping spheres on an homogeneous background \cite{Tor02,CalFra15}. The RGE for the self energy at zero momentum is very simple: it reduces to a diffussion equation with a source term. It can be solved both numerically and (under some approximations) analytically. We show the solution has an accuracy equivalent to a full two-loops self-consistent evaluation in perturbation theory.

We aim to apply this method to computing not only the total absorption from the mean wave field, as we have done here, but also the spectrum of the scattered wave, for which task not only the self energy but also the intensity operator must be known. This is a challenge to ordinary perturbation theory because, as is well known, a resummation of the simplest nonlinear approximation leads to the ladder approximation, which violates reciprocity invariance and misrepresents coherent backward scattering\cite{Knot12,KnoWel13,TigMay98,AkkMon07}. In turn, if reciprocity is restored by adding the maximally crossed graphs, then the Ward identities are no longer satisfied.

In the RG framework it is obvious that reciprocity holds at $\Lambda\to\infty$, since this is just propagation in an homogeneous medium. It must actually hold at all values of $\Lambda$, since at every value of $\Lambda$ we have propagation on a physically realizable medium. So it ought to be possible to enforce reciprocity already at the level of the RGE, which would mean a substantial improvement over a straightforward loop expansion.

We expect to report on this issue on a separate contribution.

\section*{Acknowledgements}

It is a pleasure to thank D. Barci, R. Depine, M. Franco, Z. Gonz\'alez Arenas, G. Lozano and P. Tamborenea for discussions.

This work is supported in part by CONICET and Universidad de Buenos Aires (Argentina).

\section*{Appendix: The white noise limit} 
In the white noise limit where $C(\mathbf{x},\mathbf{x'})=\eta\delta(\mathbf{x}-\mathbf{x'})$ the self energy may be obtained analytically within the bilocal (one loop) approximation\cite{Eck10}. It is interesting to see how the corresponding result is obtained from the RG.

Observe that in this limit $C\left(0\right)\to\infty$. Since $E = \e_0 + \Delta\e \, \rho v $ remains finite, this means that for white noise $e = 0$ corresponde a $C(0) \to \infty$. Concretely, the white noise limit is obtained provided 
\be 
\left\{ \begin{array}{cc}
& \Delta \e \to \infty \\
& v \to 0  \\ 
& v^2 \left(\Delta \e\right)^2 \to \frac{\eta}{\rho}
\end{array} \right.
\te 
This means that the bubble volume goes to zero but the contrast in the dielectric constant between the homogeneous medium and the bubbles diverges in such a way that the product remains finite. The momentum space correlation eq. (\ref{ckr}) becomes
\be 
\widetilde{C}_{B}\left(k \right) = \eta  
\te 
The on-shell mean value for the dielectric constant becomes 
\be 
E = \e_0  + \sqrt{\rho \eta} 
\te 
The coincidence limit of the position space correlation function diverges as $C(0) \sim \eta v^{-1}$, and so the generalized Reynolds number $Q = \w^2 R^2 \sqrt{C(0)}$ becomes
\be 
Q \sim \w^2 R^2 \sqrt{\frac{\eta}{v}} \sim \sqrt{R} \to 0 
\te 
So that ordinary perturbation theory ought to be reliable. The perturbative result is obtained neglecting the one loop contribution to the real part of the effective dielectric constant (which is formally infinite) in the right hand side of eq. (\ref{epsi1loop}). From the imaginary part of the function  $\mathcal{G}(x)$ (eq. (\ref{mathcalg})) in the limit $x\to 0$ we find 
\be 
\textrm{Im}\left(\mathcal{G}\right)(x) \sim \frac{2}{9} x + \mathcal{O}(x^2)
\te 
and so
 
\be 
\textrm{Im}(\e_1) = \frac{\eta \w^3}{4 \pi} \sqrt{E}
\label{1loopime}
\te 
We now consider this problem from the point of view of the RG. Recall the parametric form of the RGE (\ref{eex},\ref{sigme}). For $\textrm{Im}(f) = \w^{-2}R^{-2} \sigma$ we obtain 
\be 
\textrm{Im}(f) = \frac{\eta}{v \w^2 R^2} \,  \frac{3 \w^4 R^4 X^2 \left( \sin X - X \cos X\right)^2 }{X^7 + \frac{3 \w^4 R^4 \eta}{v} \left( \sin X - X \cos X\right)^2 }
\te 
and 
\be 
X^2 = \w^2 R^2 E + \frac{6}{\pi} J(X) \w^4 \eta \frac{R^4}{v} 
\te 
when $R \to 0$, $X=R\L \ll 1$ for any finite value of $\L$. Then we may approximate
\be 
\sin X - X \cos X \sim \frac{1}{3} X^3
\te 
and
\be 
X^2 \sim \w^2 R^2 E 
\te 
This yields
\be 
\textrm{Im}(f) = \frac{\eta \w^2 R^2}{3 v} \, \frac{ X }{ 1 + \frac{\eta \w^4 R^4}{3 \pi v X} } = \frac{\eta \w^3 \sqrt{E}}{4 \pi} \, \frac{1}{1 + \frac{\eta \w^3}{4\pi^2\sqrt{E}}}
\te 
in agreement with the 1-loop result eq. (\ref{1loopime}) to first order in $\eta$.

\end{document}